\documentclass[twocolumn,prb,aps,psfig,showpacs,preprintnumbers,superscriptaddress]{revtex4}
\usepackage{graphicx}
\usepackage{epstopdf}
\usepackage{float}
\usepackage{color}
\usepackage{amsmath}

\begin{document}
\title{Metal-insulator transition in antiferromagnetic  Ba$_{1-x}$K$_x$Mn$_2$As$_2$ (0 $\leq$ $x$ $\leq$ 0.4)  \\
single crystals studied by $^{55}$Mn and $^{75}$As NMR}

\author{S. Yeninas}
\affiliation{Ames Laboratory, U.S. DOE, and Department of Physics and Astronomy, Iowa State University, Ames, Iowa 50011, USA}
\author{Abhishek Pandey}
\affiliation{Ames Laboratory, U.S. DOE, and Department of Physics and Astronomy, Iowa State University, Ames, Iowa 50011, USA}
\author{V.~Ogloblichev }
\affiliation{Insutitute of Metal Physics, Ural Division of Russian Academy of Sciences, Ekaterinburg 620990, Russia}
\author{K. Mikhalev}
\affiliation{Insutitute of Metal Physics, Ural Division of Russian Academy of Sciences, Ekaterinburg 620990, Russia}
\author{D. C. Johnston}
\affiliation{Ames Laboratory, U.S. DOE, and Department of Physics and Astronomy, Iowa State University, Ames, Iowa 50011, USA}
\author{Y. Furukawa}
\affiliation{Ames Laboratory, U.S. DOE, and Department of Physics and Astronomy, Iowa State University, Ames, Iowa 50011, USA}


\date{\today}

\begin{abstract} 
    The magnetic structure and metal-insulator transition in antiferromagnetic (AFM) BaMn$_2$As$_2$ and  Ba$_{1-x}$K$_x$Mn$_2$As$_2$ single crystals have been investigated by  $^{55}$Mn and $^{75}$As nuclear magnetic resonance (NMR) measurements. 
     In the parent AFM insulator BaMn$_2$As$_2$ with a N\'eel temperature $T_{\rm  N}$  = 625 K, we observed a $^{55}$Mn zero-field NMR (ZFNMR) spectrum and confirmed the G-type AFM structure from the field dependence of the $^{55}$Mn spectra and $^{75}$As-NMR spectra below $T_{\rm N}$. 
     In hole-doped crystals with $x$ $>$ 0.01, similar $^{55}$Mn ZFNMR spectra were  observed and the AFM state was revealed to be robust up to $x$ = 0.4 with the ordered moment  nearly independent of $x$. 
      The nuclear spin-lattice relaxation rates (1/$T_1$) for both nuclei in the doped samples follow the Korringa relation $T_1T$ = const., indicating a metallic state. 
      This confirms the coexistence of AFM ordered localized Mn spins and conduction carriers from a microscopic point of view.  
      From the $x$-dependence of ($T_1T$)$^{-1/2}$ for both nuclei, we conclude that this transition is caused by vanishing of the hole concentration as the transition is approached from the metallic side.  

\end{abstract}


\maketitle

   In recent years the large family of Ba$M$$_2$As$_2$  ($M$ =  transition metal) compounds have been the subject of intensive research\cite{Johnston2010, Hellmann2007, Singh2009a}  after the discovery of the superconductivity in carrier doped BaFe$_2$As$_2$.\cite{Rotter2009} 
    Among them, BaMn$_2$As$_2$ has recently been highlighted, which exhibits an antiferromagnetic (AFM) insulating ground state\cite{Johnston2010, Singh2009a, Singh2009b, Johnston2011} and a metal-insulator transition by carrier doping\cite{Pandey2012, Bao2012} or by application of pressure.\cite{Satya2011}  
    The magnetic properties  of the parent compound BaMn$_2$As$_2$ with ThCr$_2$Si$_2$-type structure are characterized as a G-type local moment AFM with a high N\'eel temperature $T_{\rm N}$ = 625(1) K and local ordered moment $<$$\mu$$>$ = 3.88(4) $\mu_{\rm B}$/Mn at 10 K.\cite{Singh2009b}
    The moments arise from Mn$^{2+}$ ions (3$d^5$) with spin $S$ = 5/2. 
    The compound is a small-band-gap insulator with energy gap $E_{\rm gap}$ $\sim$ 0.05 eV and with a zero electronic linear heat capacity coefficient $\gamma$.\cite{Johnston2011, Singh2009a, Singh2009b} 

    Hole-doping by substitution of K for Ba results in metallic Ba$_{1-x}$K$_x$Mn$_2$As$_2$ with the same ThCr$_2$Si$_2$ crystal structure and AFM ground state.\cite{Pandey2012, Bao2012}
   This behavior is observed for $x$ as low as 1.6\%.\cite{Pandey2012} 
    The magnitude of the ordered Mn moment has been demonstrated from neutron diffraction (ND)\cite{Lamsal2013}  measurements to be nearly independent of $x$, and $T_{\rm N}$ decreases slightly with $x$ from $T_{\rm N}$ = 625 K at $x$ = 0 to 480 K at $x$ = 0.4. 
    Recent studies of single crystals report a ferromagnetic (FM) moment which coexists with the AFM local Mn moment.\cite{Bao2012, Pandey2013} 
   The FM arises below 100 K for $x$ $\geq$ 0.16 with a low temperature ($T$) ordered moment that increases to $\approx$ 0.4 $\mu_{\rm B}$/f.u. at $x$ = 0.4, attributed to half-metallic itinerant FM of the doped holes with the ordered moments aligned in the $ab$-plane.\cite{Pandey2013}
 
    In this paper, we report  $^{55}$Mn and $^{75}$As nuclear magnetic resonance (NMR) measurements to investigate the magnetic structure and metal-insulator transition in single crystals of Ba$_{1-x}$K$_x$Mn$_2$As$_2$ (0 $\leq$ $x$ $\leq$ 0.4) from a microscopic point of view. 
   Measurements of the NMR spectrum and the nuclear spin-lattice relaxation rate ($1/T_1$) provide important insight into the local spin state and conduction hole density of states at atomic sites, and into electron correlations in the system. 
    The $^{75}$As-NMR spectrum and $T$ dependence of $1/T_1$ in a powder sample of BaMn$_2$As$_2$ have been reported previously. \cite{Johnston2011}
    We will not focus on the FM observed in heavily K-doped BaMn$_2$As$_2$ in this paper. 
    The detailed studies of FM in $x$ = 0.4 compound  using magnetization, ND and NMR  have been reported recently.\cite{Pandey2013} 


    Single crystals of Ba$_{1-x}$K$_x$Mn$_2$As$_2$  were synthesized by solution growth technique either by using Sn or MnAs flux.\cite{Pandey2012, Pandey2013} 
    NMR measurements of $^{55}$Mn ($I$ = $\frac{5}{2}$; $\frac{\gamma_{\rm N}}{2\pi}$ = 10.5000 MHz/T) and $^{75}$As ($I$ = $\frac{3}{2}$; $\frac{\gamma_{\rm N}}{2\pi}$ = 7.2919 MHz/T) nuclei were conducted using a homemade phase-coherent spin-echo pulse spectrometer. 
   The $^{75}$As-NMR spectra were obtained by sweeping the magnetic field while $^{55}$Mn-NMR spectra in zero field and magnetic fields were measured in steps of frequency by either measuring the intensity of the Hahn spin-echo or taking the Fourier transform of the echo signal. 
    The pulse conditions were optimized for maximum echo intensity for each frequency point in the NMR spectra. 
    The  $1/T_1$ data  were obtained using a conventional single saturation pulse method at the central transition.
    The $1/T_1$ at each $T$ is determined by fitting the nuclear magnetization $M$ versus time $t$  using 
the multi-exponential functions $1-M(t)/M(\infty) = 0.029 e^ {-t/T_{1}} +0.18 e^ {-6t/T_{1}}  + 0.79 e^ {-15t/T_{1}}$ for $^{55}$Mn and $1-M(t)/M(\infty) = 0.1 e^ {-t/T_{1}} +0.9e^ {-6t/T_{1}}$  for $^{75}$As, \cite{Narath1967} 
    where $M(t)$ and $M(\infty)$ are the nuclear magnetization at time $t$ after the saturation and the equilibrium nuclear magnetization at $t$ $\rightarrow$ $\infty$, respectively. 

  

\begin{figure}[tb]
 \includegraphics[width=3.0in]{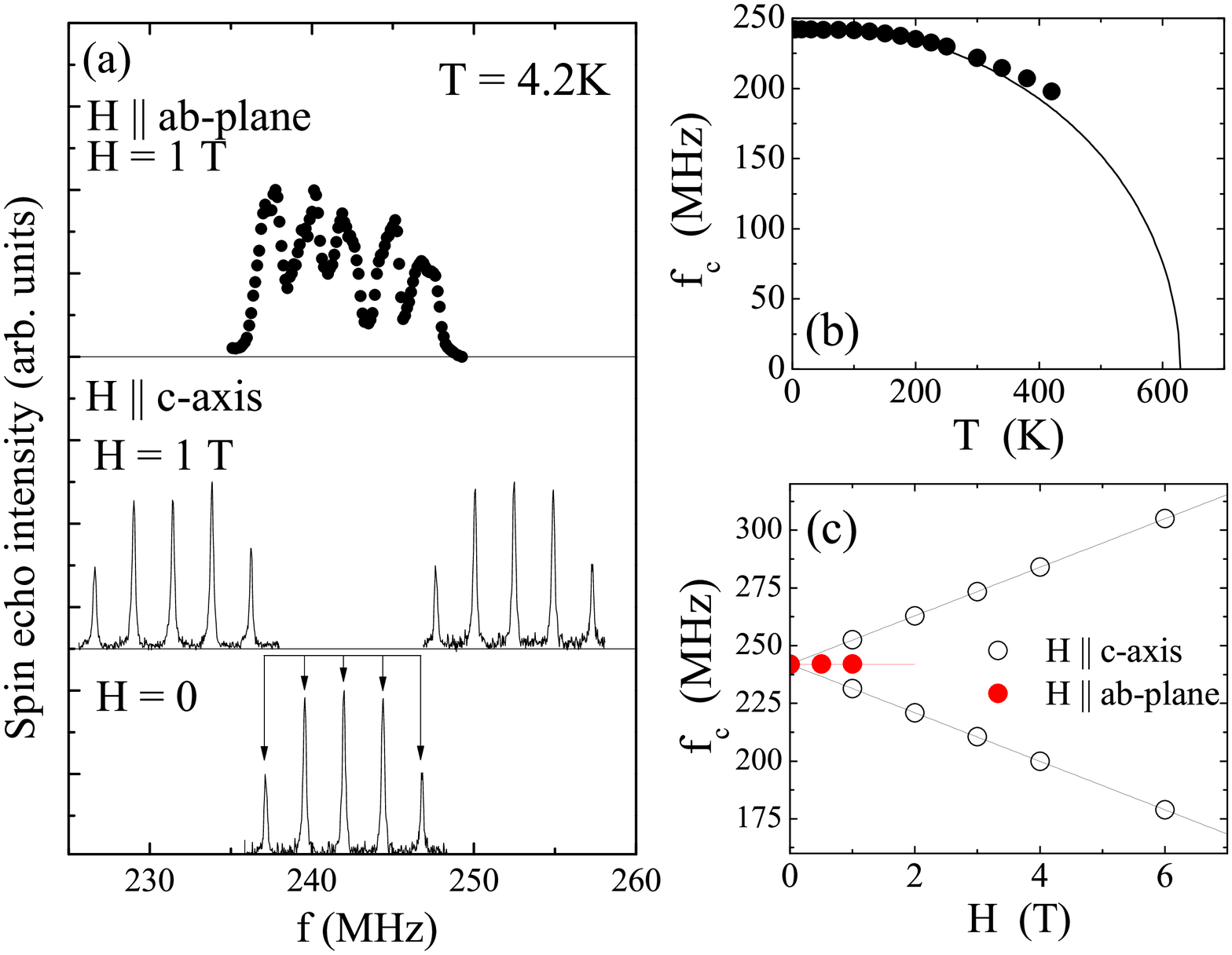}
	\caption{(Color online) (a) $^{55}$Mn-NMR spectra at $T$ = 4.2 K in the AFM state for BaMn$_2$As$_2$ in zero magnetic field (bottom), under magnetic field $H$ = 1 T parallel to the $c$-axis (middle) and perpendicular to the $c$-axis (top). 
    Arrows shown at the bottom are the calculated positions of the $^{55}$Mn NMR lines  in zero magnetic field. 
(b) Temperature dependence of the resonance frequency ($f_{\rm c}$) for the central transition line under zero magnetic field. 
     The solid line shows the $T$ dependence of $f_{\rm c}$ calculated by molecular field theory\cite {Johnston2011} with $S$ = 5/2 and $T_{\rm N}$ = 625 K.
(c) External magnetic field $H$ dependences of $f_{\rm c}$ under $H$ parallel (open circles) and perpendicular (solid circles) to the $c$-axis at $T$ = 4.2 K. }
	\label{fig:Fig.1}
\end{figure}

       At the bottom of Fig.~\ref{fig:Fig.1}(a)  the $^{55}$Mn-NMR spectrum in the AFM state for BaMn$_2$As$_2$ is shown,  measured  in zero magnetic field at temperature $T$  = 4.2 K. 
     Five sharp lines were observed, which are characteristics of nuclear spin $I$ = 5/2 with Zeeman and quadrupole interactions. 
     The sharpness of each line indicates a high quality of the single crystal. 
    The peak positions for the observed spectrum are well fitted by a second order perturbation calculation with a large internal magnetic induction $B_{\rm int}$ and a small quadrupole frequency $\nu_{\rm Q}$. 
    The vertical arrows shown in the figure are the calculated position for $^{55}$Mn zero-field NMR (ZFNMR) lines using the parameters $|B_{\rm int}|$  = 23.05(1) T, $\nu_{\rm Q}$ = 2.426(1)  MHz and $\theta$ = 0. 
Here $\theta$ represents the angle between $B_{\rm int}$ and the principle axis of the electric field gradient (EFG) tensor  at the Mn sites. 
    Since $B_{\rm int}$ is parallel to the $c$-axis as will be shown below, the principle axis of the EFG  is found to be along the $c$-axis, which is similar to the case of the Co nucleus in Ba(Fe$_{1-x}$Co$_x$)$_2$As$_2$ with the same crystal structure.\cite{Ning2009}

       $B_{\rm int}$ is proportional to $A_{\rm hf}$$<$$\mu$$>$ where $A_{\rm hf}$ is the  hyperfine coupling constant and $<$$\mu$$>$ is the ordered Mn magnetic moment. 
      The hyperfine field at the Mn sites mainly originates from core-polarization from 3$d$ electrons and is oriented in a direction opposite to that of the Mn moment. 
     For $|$$B_{\rm int}$$|$ = 23.05~T\@ and the reported AFM ordered moment $<$$\mu$$>$  = 3.88(4)~$\mu_{\rm B}$/Mn\@ from ND,\cite{Singh2009b}  $A_{\rm hf}$ is estimated to be  $-$~5.94~T/$\mu_{\rm B}$\@ where the sign is reasonably assumed to be negative due to the core-polarization mechanism. 
     This $A_{\rm hf}$ value for Mn$^{2+}$ ions  is lower than the previously reported values ranging from  $A_{\rm hf}$ = $-$~9.1~T/$\mu_{\rm B}$ in Mn-doped ZnS to $A_{\rm hf}$ = $-$~13~T/$\mu_{\rm B}$ in Mn-doped KMgF$_3$.\cite{Freeman1965}
    The difference may be explained by an additional transferred hyperfine contribution from the four nearest-neighbor Mn ions, suggesting  a large hybridization of Mn 3$d$ and As 4$p$ orbitals.
 
     The $T$ dependence of the resonance frequency ($f_{\rm c}$) for the $^{55}$Mn spectrum central transition line ($I_{z}$ = $\frac{1}{2}$$\leftrightarrow$$-\frac{1}{2}$)  shows only a slight decrease in frequency from $f_{\rm c}$ = 242.0 MHz at 4 K to 197.8 MHz at 420 K as shown in Fig.~\ref{fig:Fig.1}(b). 
  This indicates that $<$$\mu$$>$  decreases by $\sim$ 18\% from 4 K to 420 K. 
   The $T$ dependence of  $f_{\rm c}$ is similar to  the $T$ dependence of $<$$\mu$$>$ calculated with molecular field theory\cite {Johnston2011} for $S$ = 5/2 and $T_{\rm N}$ = 625 K, as shown by the solid line.

      In order to determine the direction of $B_{\rm int}$ with respect to the crystal axes, 
we measured $^{55}$Mn NMR in a single crystal under an external magnetic field $H$. 
   When $H$ is applied along the $c$-axis, each line splits into two lines as shown in the middle panel of Fig.~\ref{fig:Fig.1}(a).  
     The $H$ dependence of $f_{\rm c}$ is shown in Fig.~\ref{fig:Fig.1}(c) and the slopes of the field dependence of $f_{\rm c}$ for both lines are  $\pm$10.5 MHz/T, which is exactly the same as $\frac{\gamma_{\rm N}}{2\pi}$ of the $^{55}$Mn nucleus.   
    Since the effective field at the Mn site is given by the vector sum of  $\vec{B}_{\rm int}$ and $\vec{H}$, i.e., $|$$\vec{B}_{\rm eff}$$|$ = $|$$\vec{B}_{\rm int}$ + $\vec{H}$$|$,   the resonance frequency is expressed as $f$ = $\frac{\gamma_{\rm N}}{2\pi}$$|\vec{B}_{\rm eff}|$. 
   Thus the $H$ dependence of the spectra clearly indicates that the Mn magnetic moments for each of the two sublattices in the AFM state are parallel or antiparallel to the $c$-axis.        

     In the case of $H$ applied perpendicular to the $c$-axis, no splitting of the $^{55}$Mn-NMR lines is observed (see top panel in Fig.~\ref{fig:Fig.1}(a)).  
    In this orientation the applied field is orthogonal to the ordered Mn moments and thus to $B_{\rm int}$, so one expects $f$ =  $\frac{\gamma_{\rm N}}{2\pi}$$\sqrt {B_{\rm int}^2+H^2}$ .      
    For our applied field range, $B_{\rm int}$ $>>$ $H$, any shift in the resonance frequency would be small, which is observed as shown by solid circles in  Fig.~\ref{fig:Fig.1}(c). 
   These results are consistent with the G-type AFM spin structure reported from ND measurements with the ordered moments aligned along the $c$-axis.  
   The observed broadening of each line is likely caused by a small misalignment between the $H$ and the $ab$-plane, which introduces a small component of the external magnetic field along the $c$-axis. 
    From the observed broadness, the misalignment is estimated to be $\sim$3$^{\circ}$.  
   
\begin{figure}[tb]
  \includegraphics[width=3.0in]{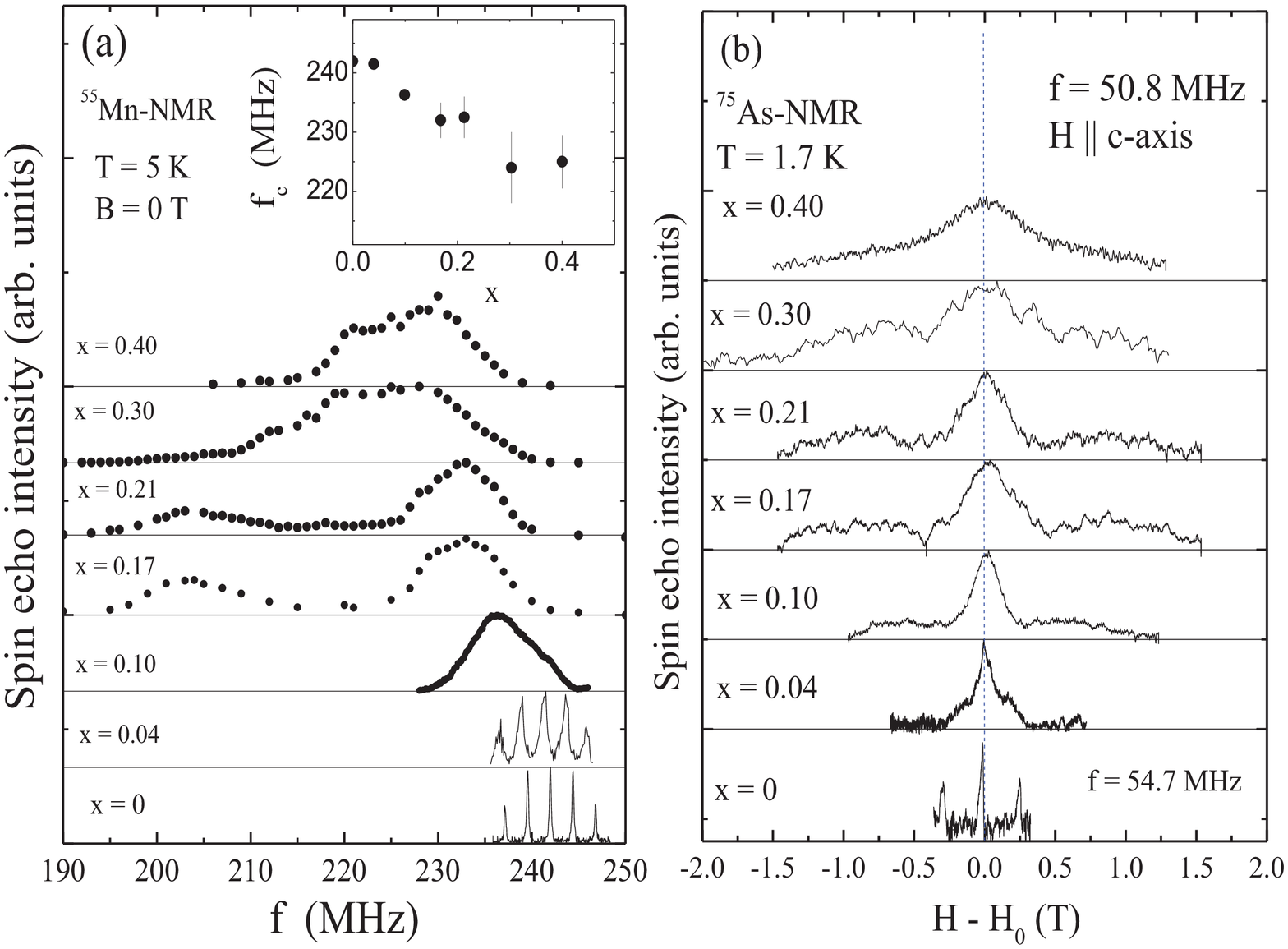}
	 \caption{(Color online) (a) $^{55}$Mn zero-field NMR spectra in Ba$_{1-x}$K$_x$Mn$_2$As$_2$  at $T$ = 5 K.   
      Inset: $x$ dependence of $f_{\rm c}$ at $T$ = 5 K.
       (b) Field-swept $^{75}$As-NMR spectra in Ba$_{1-x}$K$_x$Mn$_2$As$_2$. 
          For $x$ = 0, the spectrum was measured at $f$ = 54.7 MHz at $T$ = 75 K. 
          For $x$ $\geq$ 0.04, the spectra were measured at $f$ = 50.8 MHz with $H$ parallel to the $c$-axis. The horizontal axis is shifted by the corresponding Lamor field $H_0$. } 
	\label{fig:Fig.2}
\end{figure}

      $^{55}$Mn ZFNMR spectra for K-doped crystals show an increase of the line widths and a shift to lower frequency upon increasing $x$ as shown in Fig.~\ref{fig:Fig.2}(a). 
    The broadening of the lines is attributed to an increased internal field distribution due to Mn  moment distribution and an increased $\nu_{\rm Q}$ distribution due to local lattice distortion as a result of K doping. 
    While five lines are observed for $x$ = 0.04, the line widths for $x$ $\geq$ 0.10 become too large to observe distinct quadrupole-split lines. 
    For $x$ $\geq$ 0.17, an enhancement of exciting radio-frequency field $H_1$, characteristic of ferromagnetic substances,\cite{Portis1965} is observed at $T$ = 5 K and is estimated to be $\sim$ 60. 
    This indicates the coexistence of FM ordering with the G-type AFM structure in Ba$_{1-x}$K$_x$Mn$_2$As$_2$ for $x$ $\geq$ 0.17, consistent with recent studies.\cite{Bao2012, Pandey2013} 

    Secondary broad NMR signals around 200 MHz for intermediate concentrations $x$ =0.17 and 0.21 were observed as shown in Fig.~\ref{fig:Fig.2}(a), whose origin is not clear at present.  
     A possible origin is $^{75}$As-NMR from a FM MnAs impurity phase.\cite{Pinjar1982}
    If this signal indeed arises from MnAs, then one would expect a $^{55}$Mn signal at 231 MHz $<$ $f$ $<$ 239 MHz from the impurity phase\cite{Pinjar1982, Hihara1962}  which would overlap with the intrinsic signal. 
     Therefore, since the signal intensity from the impurity seems to be relatively large for $x$ = 0.17, we use our $^{55}$Mn-NMR data for $x$ = 0.17 only for reference.  

     The inset of Fig.~\ref{fig:Fig.2}(a)  plots the center frequency $f_{\rm c}$ of the $^{55}$Mn spectra as a function of $x$  where we determined $f_{\rm c}$ as the areal median of the experimental spectrum  for $x$ $\geq$ 0.1. 
    With increasing $x$, the spectrum slightly shifts to lower frequencies, indicating a decrease in the local hyperfine field. 
     If one assumes the same hyperfine coupling constant $A_{\rm hf}$ for the K-doped system as for the undoped compound, this indicates a decrease of only $\sim$ 8\% in the average Mn ordered  moment from $x$ = 0 to $x$ = 0.4. 
     This would be consistent  with results from ND which report almost no change in the Mn moment from $<$$\mu$$>$ = 3.88(4) $\mu_{\rm B}$/Mn for $x$ = 0 to 3.85(15) $\mu_{\rm B}$/Mn   for $x$ = 0.4.\cite{Lamsal2013}

     In order to investigate the Mn AFM structure in K-doped single crystals, we measured $^{55}$Mn-NMR spectra under magnetic fields applied parallel to the $c$-axis and to the $ab$-plane as in the case of the undoped BaMn$_2$As$_2$.    
    The observed field dependence of the  $^{55}$Mn-NMR spectra for the K-doped crystals was similar to BaMn$_2$As$_2$, which indicates the occurrence of the same AFM Mn spin structure in K-doped crystals. 
      These results are consistent with the G-type AFM ordering in Ba$_{1-x}$K$_x$Mn$_2$As$_2$ from $x$ = 0 to $x$ = 0.4,  as previously reported.\cite{Lamsal2013}

    Figure~\ref{fig:Fig.2}(b) shows field-swept $^{75}$As-NMR spectra for Ba$_{1-x}$K$_x$Mn$_2$As$_2$ with 0.04 $\leq$ $x$ $\leq$ 0.40. 
     Here, for comparison,  we also show the previously reported $^{75}$As-NMR spectrum for $x$ = 0 which shows a clear quadrupole splitting for $I$ = 3/2. \cite{Pandey2013} 
    The quadrupole frequency $\nu_{\rm Q}$ = 2.01 MHz of the $^{75}$As nucleus is nearly $T$-independent in our observed temperature range of $T$ = 4.2 -- 300~K\@. 
    The central transition peak lies just below the unshifted Larmor field $H_0$ denoted by the vertical dashed line,  suggesting that the average internal field at the As sites is approximately zero.  
    The zero internal field at the As sites in the AFM state is consistent with the G-type AFM ordering in BaMn$_2$As$_2$, as has been pointed out by $^{75}$As NMR spectrum measurements on single crystal\cite{Pandey2013} and polycrystalline\cite{Johnston2011}  samples.

     With K-doping, the observed spectra shown in Fig.~\ref{fig:Fig.2}(b)  become broad and  the sharp quadrupole peaks are smeared out.
    The broadening increases with increasing $x$ and the signal intensity is reduced, requiring measurements be taken at low temperatures. 
    Despite the broad spectra, the average internal field is approximately zero, as observed in the undoped system. 
    The lack of any observed internal field at the $^{75}$As site supports a local-moment G-type AFM ordering for K-doped compounds.\cite{Lamsal2013}

\begin{figure}[tb]
   \includegraphics[width=3.0in]{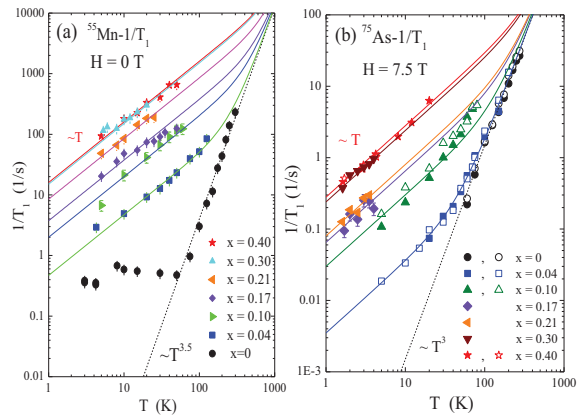}
     \caption{(Color online) (a) Temperature dependence of $^{55}$Mn-$1/T_1$ at $H$ = 0 in Ba$_{1-x}$K$_x$Mn$_2$As$_2$ (0$\leq$ $x$ $\leq$ 0.40). 
        The dotted line shows 1/$T_1$ $\propto$ $T^{3.5}$ power-law behavior for $x$ = 0 and solid lines represent the Korringa relation ($T_1T$) = const for $x$ $\geq$ 0.04.  
        (b)  Temperature dependence of $^{75}$As-$1/T_1$ at $H$ = 7.5 T.  
     Solid and open symbols are for $H$ $||$ $c$-axis and $H$ $||$ $ab$-plane, respectively. 
       The dotted line shows 1/$T_1$ $\propto$ $T^{3}$ power-law behavior for $x$ = 0 and the solid lines represent the Korringa relation $T_1T$ = const. } 
              \label{fig:Fig.3}
\end{figure}


   We now discuss the dynamical properties of the electronic and magnetic states, based on 1/$T_1$ data measured for $^{55}$Mn and $^{75}$As nuclei.
    Figures~\ref{fig:Fig.3}(a)  and \ref{fig:Fig.3}(b) show the $T$-dependences of $1/T_1$  for $^{55}$Mn in $H$ = 0 and for $^{75}$As in $H$ = 7.5 T, respectively. 
     In the latter figure, $1/T_1$  of $^{75}$As measured with $H$~$||$~$c$-axis  and $H$~$||$~$ab$-plane are shown by solid  and open symbols, respectively.

   In the case of the BaMn$_2$As$_2$, $1/T_1$ of both $^{55}$Mn and $^{75}$As shows a strong $T$ dependence for $T$ $>$ 40 K where  1/$T_1$ shows $T^{3.5 \pm 0.3}$ and $T^{3.0 \pm 0.2}$ power-law behaviors (shown by the dotted lines in the figure) for $^{55}$Mn and $^{75}$As, respectively.   
     This power-law $T$-dependence for both 1/$T_1$ measurements can be explained by a two-magnon Raman process as the main relaxation mechanism for an AFM insulating state when $T$ $>>$ $\Delta$, where $\Delta$ is the anisotropy gap energy in the spin wave spectrum.\cite{Beeman1968}  
      The $T^3$ dependence of the $^{75}$As-1/$T_1$ in the AFM state in polycrystalline BaMn$_2$As$_2$ samples was previously reported.\cite{Johnston2011}
   The deviation from the power-law behavior for the $^{55}$Mn-1/$T_1$  below 40 K is likely due to relaxation associated with impurities. 

     In the hole-doped samples, 1/$T_1$ of both $^{55}$Mn and $^{75}$As show a Korringa relation $(T_1T)^{-1}$ = const  at low $T$, demonstrating direct evidence of conduction electrons at the Mn and As sites. 
    This confirms that hole-doping, even at our lowest observed concentration $x$ = 0.04, results in a metallic ground state for the Ba$_{1-x}$K$_x$Mn$_2$As$_2$ system.

\begin{figure}[tb]
  \includegraphics[width=3.5in]{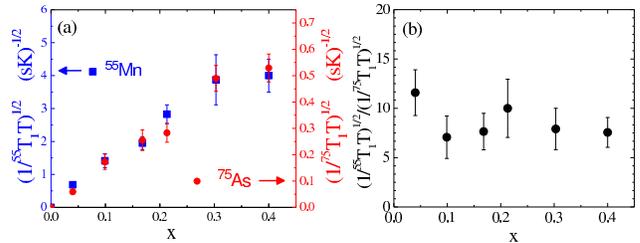}
	 \caption{(Color online) (a) $x$ dependence of ($T_1T$)$^{-1/2}$ for both $^{55}$Mn and $^{75}$As nuclei. 
           (b) $x$ dependence of ratio of ($T_1T$)$^{-1/2}$ for $^{55}$Mn and $^{75}$As nuclei. 
} 
	\label{fig:Fig.4}
\end{figure}

   The $T$-independent  $1/T_1T$ can be expressed in terms of the density of states ${\cal D}(E_{\rm F})$ at the Fermi level as\cite{Narath_Hyper} $(T_{1}T)^{-1}=4\pi\gamma_{\rm N}^2\hbar k_{\rm B}A_{\rm hf}^2
{\cal D}^2(E_{\rm F})$  where $k_{\rm B}$  is Boltzmann's constant. 
    As shown in Fig.~\ref {fig:Fig.4}(a),  ($T_1T)^{-1/2}$  
for both $^{55}$Mn and $^{75}$As nuclei increase with $x$, indicating an increase in the local ${\cal D}(E_{\rm F})$  at each atomic site upon increasing $x$. 
    When plotted as a ratio of  ($T_1T)^{-1/2}$ for $^{55}$Mn and $^{75}$As in Fig.~\ref{fig:Fig.4}(b), it is noticed that the ratio is almost constant over our range of $x$ within our experimental uncertainty.  
    This means that the ratio of ${\cal D}(E_{\rm F})$ at the Mn sites to ${\cal D}(E_{\rm F})$ at the As sites is independent of $x$, suggesting a rigidity of band structure upon K-substitution.
    According to the band calculations with the rigid band model,\cite{Pandey2012} the total ${\cal D}(E_{\rm F})$ consists of  66(60)\%  Mn-3$d$ and 28(30)\% As-4$p$ bands for $x$ = 0.016(0.05) 
and the ratio of ${\cal D}(E_{\rm F})$ for the Mn and As bands is almost independent of $x$.
    Thus our 1/$T_1$ data  seem to be consistent with the band calculations.

    Finally we discuss a plausible mechanism of the metal-insulator transition for the present system. 
    Electrical conductivity is expressed as $\sigma$ =  $n_0$$e^2$$\tau$/$m^*$  where $n_0$ is the carrier density,  $m^*$  is effective carrier mass and $\tau$ is relaxation time, therefore  a transition from metal to insulator can be characterized by $n_0$ $\rightarrow$ 0 or $m^*$ $\rightarrow$ $\infty$ or $\tau$ $\rightarrow$ 0. 
     In a typical case of strong correlated  electron systems, for example La$_{1-x}$Sr$_x$TiO$_3$ with perovskite structure, the Mott insulator–-metal transition is characterized by a divergent behavior of $m^*$ due to strong electron correlation effects,\cite{Kumagai1993} where the $T$-independent  $(T_1T)^{-1}$ of $^{47/49}$Ti and $^{139}$La show a significant enhancement as the transition is approached from the metallic side.\cite{Furukawa1999}
     On the other hand, in the Ba$_{1-x}$K$_x$Mn$_2$As$_2$ system, the mechanism of the metal-insulator transition is evidently different. 
       The decrease of ($T_1T$)$^{-1}$ for both $^{55}$Mn and $^{75}$As nuclei with decreasing $x$  indicates that the local density of states at both sites decrease with decreasing $x$.
     Thus we conclude that the transition from metal to insulator in the present system is characterized by the decrease of $n_0$. 
    The details of the carrier doping effects in Ba$_{1-x}$K$_x$Mn$_2$As$_2$ are still open question. 
     BaMn$_2$As$_2$ is considered to be an AFM insulator where five electrons with parallel spins on Mn 3$d$ orbitals are localized due to strong electron correlations. 
     Thus one may expect to observe electron correlation effects in the metallic state near the boundary at the metal-insulator phase transition. 
    However, we have not observed any clear sign of the effects of strong electron correlations near the phase boundary from the present $1/T_1T$ data. 
     In contras,  the importance of the electron correlation effects has been pointed out in resistivity and specific heat measurements.\cite{Pandey2012}
     Since $1/T_1$ measurements prove dynamical properties of conduction electrons at the NMR frequency  ($\omega$ $\sim$ 10 -- 500 MHz) which is contrast to the static measurements ($\omega$ $\sim$ 0) such as the resistivity and the specific heat measurements, the discrepancy may suggest that  the effects of the electron correlations in Ba$_{1-x}$K$_x$Mn$_2$As$_2$ depend on frequency of dynamical properties of the electrons and enhance at $\omega$ $\sim$ 0. 
     We also observed a $H_1$ enhancement effect which supports recent studies for coexistence of AFM ordered Mn-3$d$ moments and FM for $x$ $\geq$ 0.16 where the  mechanism of the FM is not still clear.  
   Further studies from both experimental  and theoretical sides are required to understand the details of carrier doping effects in Ba$_{1-x}$K$_x$Mn$_2$As$_2$ in which the novel half-metallic FM ordering appears  in $x$ $\geq$ 0.16.\cite{Bao2012, Pandey2013}

       In summary, we report $^{55}$Mn and $^{75}$As NMR results for insulating BaMn$_2$As$_2$ and hole-doped metallic Ba$_{1-x}$K$_x$Mn$_2$As$_2$ ($x$  = 0.04 -- 0.4) single crystals. 
   $^{55}$Mn and $^{75}$As NMR spectrum measurements confirm similar G-type local-moment AFM structures for both insulating and metallic states. 
    $^{55}$Mn spectra suggest that the local Mn ordered magnetic  moments are robust and are almost independent of hole doping, consistent with the ND results. 
    The $T$-dependence of $1/T_1$ for $^{55}$Mn and $^{75}$As NMR confirm an insulating ground state for BaMn$_2$As$_2$ and metallic ground states in hole-doped Ba$_{1-x}$K$_x$Mn$_2$As$_2$  from a microscopic point of view, evidence of metal-insulator transition in AFM Ba$_{1-x}$K$_x$Mn$_2$As$_2$. 
   The metal-insulator transition is revealed to be characterized by vanishing of the carrier (hole) concentration as the transition is approached from the metallic side.

    This research was supported by the U.S. Department of Energy, Office of Basic Energy Sciences, Division of Materials Sciences and Engineering. Ames Laboratory is operated for the U.S. Department of Energy by Iowa State University under Contract No.~DE-AC02-07CH11358.


\end{document}